\documentclass{article}
\usepackage{graphicx}
\usepackage{amsfonts}
\usepackage{amsmath}
\usepackage{hyperref}
\usepackage{comment}
\usepackage[english]{babel}

\begin{document}

\begin{center}
    \LARGE \textbf{Copula-based deviation measure of cointegrated financial assets}
\end{center}

\vspace{10mm}

\begin{center}
    \Large Alexander Shulzhenko \\
    \vspace{5mm}
    \normalsize Faculty of Math and Computer Science, \\
    Saint-Petersburg State University \\
    \vspace{5mm}
    \texttt{shulzhenko2426@gmail.com}
\end{center}

\vspace{10mm}

\begin{abstract}
    This study outlines a comprehensive methodology utilizing copulas to discern inconsistencies in the behavior exhibited by pairs of financial assets. It introduces a robust approach to establishing the interrelationship between the returns of these assets, exploring potential measures of dependence among the stochastic variables represented by these returns. Special emphasis is placed on scrutinizing the traditional measure of dependence, namely the correlation coefficient, delineating its limitations. Furthermore, the study articulates an alternative methodology that offers enhanced stability and informativeness in appraising the relationship between financial instrument returns.
\end{abstract}


\section{Introduction}
The study of the relationship between financial assets in particular and financial markets in general today remains one of the central problems in financial mathematics and econometrics. Measuring the relationship between financial assets underlies fundamental financial theories such as: Capital Asset Pricing Theory \cite{CAPM} and Arbitrage Pricing Theory \cite{APT}. 

Investors exhibit a keen interest in effectively and resiliently estimating dependencies among diverse financial instruments, as this estimation facilitates the application of \textit{statistical arbitrage}. Contemporary discourse encompasses a comprehensive theoretical framework on ``pair trading'' \cite{Gatev}, wherein a number of methodologies is employed to gauge the interrelationship between securities.

The overarching concept of "pair trading" is relatively straightforward. Conceptually, the methodology can be divided into two primary steps:
\begin{enumerate}
    \item Identify two securities whose returns historically moved "in the same direction";
    \item Monitor the change in spread during the trading period between the chosen securities. Assuming a statistically significant increase in spread, under the assumption of maintaining historical correlation, the returns of the selected assets will revert to historical equilibrium. This allows for profit extraction by taking a long (short) position on the "underpriced" ("overpriced") asset.
\end{enumerate}

The description of the ``pair trading'' method illustrates that a pivotal role in this approach is played by the method of measuring the ``interaction'' between securities, as well as the method for measuring the significance of the deviation in spread from historical values. A classical method for measuring the dependence between two securities is the so-called ``distance method'', intricately detailed in the paper by Gatev et al. \cite{Gatev}. Another equally popular method for measuring dependence involves computing the Pearson correlation coefficient\footnote{It's worth noting that the concept of correlation holds a central place in financial theory overall. Risk management theory, as well as the mathematical methodology underlying insurance, utilizes correlation to measure the statistical relationship between risks.} \cite{Chen}. In more detail, correlation coefficients between the returns of securities are determined, after which the authors compute the divergence metric $D_{ijt}$:
$$ D_{ijt} = \beta (r_{it} - r_f) - (r_{jt} - r_f)$$,
where $\beta$ stands for the regression coefficient of returns $r_{it}$ of security $i$ on returns $r_{jt}$ of security $j$, $r_f$ is the risk free rate. Chen et al. \cite{Chen} investigate two possibilities of choosing $r_{jt}$:
\begin{enumerate}
    \item \textit{Univariate}: $r_{jt}$ -- returns of the most correlated security (with returns of the security $i$);
    \item \textit{Quasi-multivariate}\footnote{The generalization of this approach is often refereed as ``Portfolio management theory''}: $r_{jt}$ -- returns of the portfolio containing $50$ securities (uniformly weighted) with the highest correlations with the returns of the security $i$.
\end{enumerate}

Correlation, being one of the central concepts in the ``pairs trading'' methodology, has a number of disadvantages, which are described in detail in the next section. 

\section{Correlation as a measure of dependence} 
\subsection{Distribution of returns} \label{fin_distr}
Correlation serves as the canonical measure of dependence for multivariate normal distributions, or more generally, for elliptical distributions. However, distributions obtained from the analysis of financial data typically poorly conform to the distributions of these mentioned classes. To be more precise, these distributions exhibit much heavier tails than the exponentially decreasing tails of the Gaussian distribution.

For example, Ibragimov, Prokhorov \cite{Ibragimov_ht}, Ibragimov et al. \cite{Ibragimov_new_appr}, consider \textit{power law family} distributions, namely, distributions with tails decreasing as inversed polynomials:
$$ \mathbb{P}(|X| > x) \sim x^{-\alpha}.$$

It is worth noting an important fact that the $p$-th absolute moment of a random variable $X$ belonging to power law family is finite if and only if its order is less than index $\alpha$:

\begin{equation}
\label{eqn:tails_prop}
    p < \alpha \implies \mathbb{E}|X|^p < \infty; \quad  p \ge \alpha \implies \mathbb{E}|X|^p = \infty
\end{equation}

The above observations are summarized by the so-called ``stylized facts'', which describe the empirical properties of asset return distributions \cite{stylized}.

\subsection{Specifics of using correlation as a measure of dependence}

\textbf{Definition.} Number $\rho$ is called the linear correlation coefficient between random variables $X$ and $Y$:

$$ \rho(X, Y) = \frac{\mathbf{Cov}[X, Y]}{\sqrt{\sigma^2[X] \sigma^2[Y]}}$$,
where $\mathbf{Cov}[X, Y]$ -- covariance between $X$ and $Y$, $\sigma^2[X]$, $\sigma^2[Y]$ -- variances of $X$ and $Y$ respectively. 

As noted earlier, correlation stands as an immensely popular tool when working with multivariate normal distributions (or more generally, with multivariate spherical and elliptical distributions) since it serves as a natural measure of dependence within this class of distributions. Specifically, an elliptical distribution is uniquely determined by its mean and covariance matrix.

The properties of correlation for elliptical distributions, however, do not extend to many other classes. Consequently, the correlation coefficient as a measure of dependence between random variables has certain significant drawbacks \cite{pitfalls}. The most notable ones are listed below:
\begin{enumerate}
    \item The variances of $X$ and $Y$ must be finite; otherwise, the correlation coefficient is undefined. This property poses challenges when working with heavy-tailed distributions (see property \ref{eqn:tails_prop}), which, as noted, often emerge in modeling returns.
    \item Independence between random variables implies that they are uncorrelated ($\rho=0$); however, the converse is generally not true. 
    \item Correlation coefficient is not invariant under the non-linear strictly monotonically increasing transforms: 
    \begin{equation}
        \rho(X, Y) \neq \rho(T(X), T(Y)), \quad T: \mathbb{R} \to \mathbb{R}.
    \end{equation}
    \item Marginal distributions of $X$ and $Y$ and the correlation coefficient do not determine uniquely join distribution of the random vector $(X, Y)$. 
\end{enumerate}

\begin{figure}[!h]
    \centering
    \includegraphics[width=70mm]{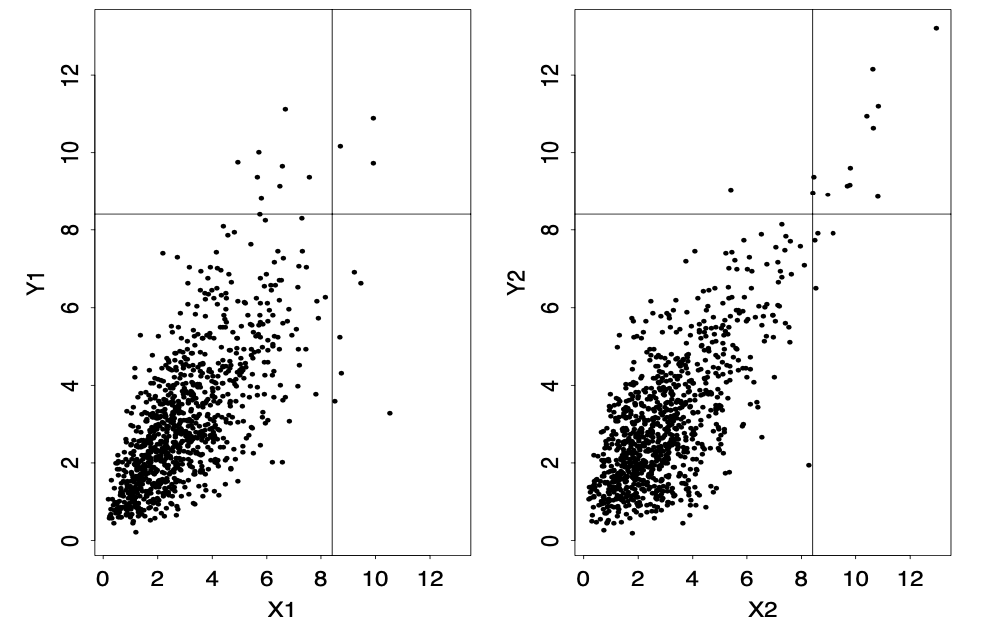}
    \caption{$1000$ samples of random vectors with equal marginal distributions and equal correlation coefficients, however with different joint distributions.}
    \label{fig:gg}
\end{figure}

Point $4$ explicates the fallacy that the joint distribution is fully characterized solely by the marginal distributions and the correlation coefficient. Figure \ref{fig:gg} illustrates examples of random vectors having identical marginal distributions and equal correlation coefficients between their components but exhibiting different dependency structures.

The preceding argument indicates that while the correlation coefficient is a popular metric in ``pairs trading'', it is unable to fully describe the relationship between two random variables. This limitation potentially leads to an inefficient assessment of the interaction between two assets returns. This observation gains particular relevance in light of the discussions in Section \ref{fin_distr}. 

A natural question is to find a tool that would allow us to comprehensively describe the structure of the dependence of two random variables based on their marginal distributions.

\section{Copulas: definitions and basic properties}
\subsection{General information} \label{copulas_basics}

Dependence structure between $n$ real-valued random variables $X_1, \dots, X_n$ is completely determined by their joint probability function: 
$$ F(x_1, \dots, x_n) = \mathbb{P}[X_1 \le x_1, \dots, X_n \le x_n].$$

The concept of copula allows to divide the function $F$ into parts, one of which describes the structure of the dependence, and the others only marginal distributions of random variables $X_i$.

Consider $n$ dimensional random vector $\textbf{X} = (X_1, \dots, X_n)^T$ with continuous marginal distribution functions $F_1, \dots, F_n$. Applying transformation\footnote{Probability integral transformation}:
$$ T \colon \mathbb{R}^n \to \mathbb{R}^n, \; (x_1, \dots, x_n) \mapsto (F_1(x_1), \dots, F_n(x_n))$$
element-wise to the components of vector \textbf{X}, we will get that all the marginal distributions of the given vector are uniform on $[0, 1]$ ($ T \textbf{X}_i \sim U(0, 1)$). The joint distribution function $C$ of the vector $(F_1(x_1), \dots, F_n(x_n))^T$ is called \textbf{copula} of a random vector \textbf{X}:
$$ F(x_1, \dots, x_n) = \mathbb{P}[F_1(X_1) \le x_1, \dots, F_n(X_n) \le x_n] = C(F_1(x_1), \dots, F_n(x_n)).$$

\textbf{Definition.} A \textbf{copula} is a distribution function of a random vector in $\mathbb{R}^n$ with uniform marginal distributions of components on $[0, 1]$.

In other words, a copula is a function $C \colon [0, 1]^n \to [0, 1]$ that has three properties:

\begin{enumerate}
    \item $C(x_1, \dots, x_n)$ is strictly increasing in each argument $x_i$;
    \item $C(1, \dots, 1, x_i, 1, \dots, 1) = x_i$ for all $i$;
    \item For all $(a_1, \dots, a_n)$, $(b_1, \dots, b_n) \in [0, 1]^n$ and $a_i \le b_i$:
    $$ \sum^2_{i_1 = 0} \dots \sum^2_{i_n = 0} (-1)^{i_1 + \dots + i_n} C(x_{i_1}, \dots, x_{i_n}) \ge 0 $$
\end{enumerate}

Thus, copula allows to describe the joint distribution of random variables using their marginal distributions \cite{sklar}. 

\subsection{Properties of copulas} \label{prop}

Copulas have the properties that were mentioned in the section $2.2$, namely:

\begin{enumerate}
    \item If the marginal distribution functions $F_1, \dots, F_n$ are continuous, then the copula $C$ is uniquely defined and
    $$ C(u_1, \dots, u_n) = F_{X_1, \dots, X_n}(F^{-1}_1(u_1), \dots, F^{-1}_n(u_n)), $$
    where $F^{-1}(u) = \textbf{inf}\{ x \colon F(x) \ge u \}$.
    \item If $X$ and $Y$ are independent r.v., then $C(u, v) = uv$.
    \item R.v. $T(X)$ and $T(Y)$ have the same copula as $X$, $Y$, where $T(\cdot)$ is a strictly monotonically increasing transformation (property $3$ for the correlation coefficient).
    \item Conditional distribution is described by the first derivative of the copula: 
    $$ F_{Y \mid X}(y) = \mathbb{P}[Y \le y \mid X = x] = \frac{\partial}{\partial u} C(u, v) $$
    
\end{enumerate}

\section{Copulas in pairs trading}

Copulas represent a relatively recent tool in pair trading methodology, yet the copula approach possesses a rigorous mathematical foundation and has been studied extensively in works by Krauss \& Stubinger \cite{krauss_stub}, Ferreira \cite{ferreira}, Liew \& Wu \cite{liew_wu}, Stander et al. \cite{stander}, and Keshavarz et al. \cite{keshavarz}. These studies employed point estimations of return deviations from historical values utilizing partial derivatives of copulas. Meanwhile, Xie et al. \cite{xie}, Rad et al. \cite{rad}, and Silva et al. \cite{silva} proposed employing the Copula Mispricing Index (CMI\footnote{Copula Mispricing Index}), which, beyond current values, also integrates previous observations.

The methodology described in this work generalizes the results obtained in the above-mentioned works, and also uses the latest results obtained by Tadi, Witzany \cite{czech}. The methodology consists of the following key steps:

\begin{enumerate}
    \item \textbf{Define the spread}. In this study, adhering to the methodology outlined in the research by Tadi \& Witzany \cite{czech}, we denote by spread a random process:
    $$ S^i_t = BASE_t - \hat{\beta}^i P^i_t, $$
    where $BASE_t$ -- returns of the base asset, which is chosen specifically for each market\footnote{For crypto currencies market, for example, natural choice is \texttt{BTCUSDT}}, and $P^i_t$ -- returns of the trading asset, which will be used in the trading pair. 
    \item \textbf{Determine trading pairs}. The selection of suitable pairs relies on historical data and statistical tests, such as the Engle-Granger test and nonlinear cointegration tests. The selection of trading pairs does not constitute the primary focus of this study; for a more detailed methodology on conducting such tests, it is outlined, for instance, in the research by Leung et al. \cite{coint}.
    \item \textbf{Estimate marginal distribution parameters}. At this stage, we assume that the trading pair has already been identified, and consequently, the historical returns of each of the two assets constituting the trading pair are known. To construct the copula, it is necessary to establish the form of the marginal distributions, for which various parametric distribution families undergo testing. Natural choices for these families include: 
    \begin{itemize}
        \item Gaussian random variables: $X \sim \mathcal{N}(\mu, \sigma)$;
        \item Student random variables: $X \sim t(n)$;
        \item Cauchy random variables: $X \sim \mathcal{C}(x_0, \gamma)$.
    \end{itemize}

    Probability density functions of these distributions are presented on the fig. \ref{fig:pdfs}.

    \begin{figure}[!h]
        \centering
        \includegraphics[width=90mm]{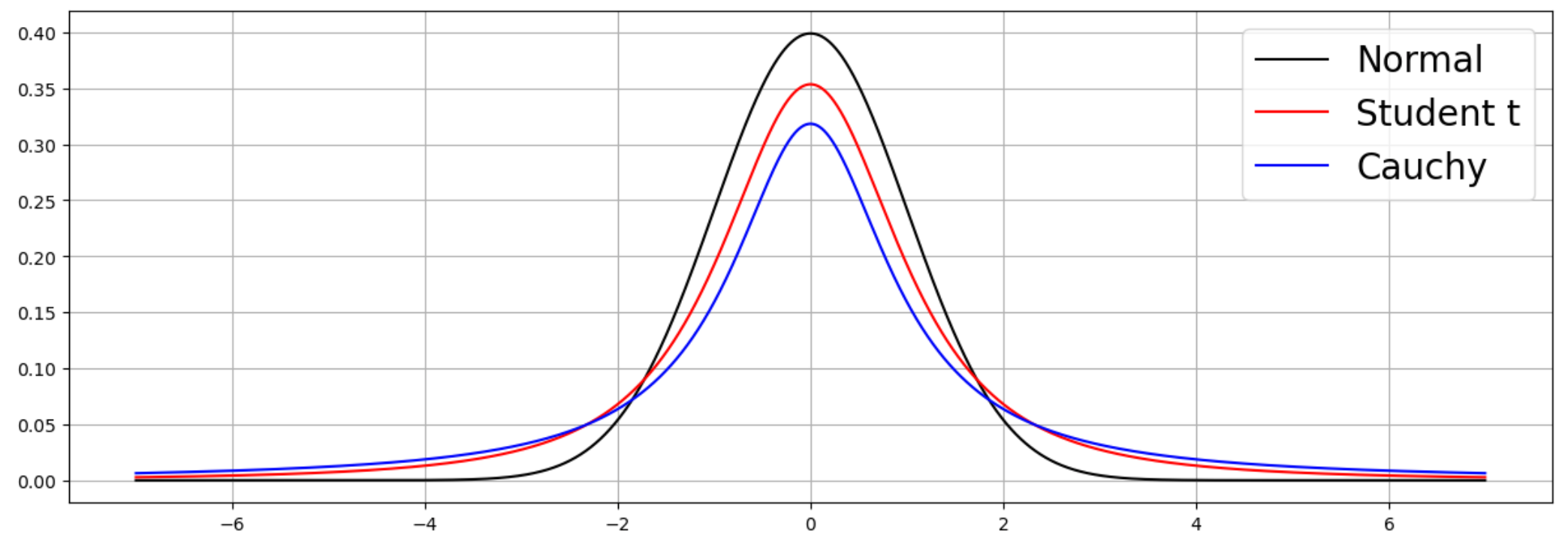}
        \caption{Probability density functions of Gaussian r.v., Student r.v. and Cauchy r.v.}
        \label{fig:pdfs}
    \end{figure}
    
    Using the maximum likelihood method, one can obtain an estimate of the distribution parameters for each of the mentioned classes, after which the ``quality'' of the model can be determined, for example, using the Akaike test (AIC).
    \item \textbf{Estimate copula parameters}. At this stage, we have obtained the explicit form of the marginal distribution functions $F^i$ for spreads $S^i_t$. By employing the integral transformation detailed in Section \ref{copulas_basics}, we acquire random variables $U^i = F^i(S^i_t)$ with uniform distributions. Subsequently, we need to estimate the copula parameters for the random variables $U^i$. Similar to the consideration of marginal distributions, various classes of copulas are scrutinized. The most commonly used are:
    
    \begin{itemize}
        \item \textbf{Independent}. $$C^{Ind}(u, v) = uv;$$
        \item \textbf{Clayton}. $$C^{Clayton}(u, v) = \left(u^{-\theta} + v^{-\theta} - 1 \right)^{-\frac{1}{\theta}};$$
        \item \textbf{Gumbel}. $$C^{Gumbel}(u, v) = \textbf{exp}\left( -[(-\textbf{ln} \: u)^{\theta} + (-\textbf{ln} \: v)^{\theta}]\right)^{-\frac{1}{\theta}}; $$
        \item \textbf{Eyraud-Farlie-Gumbel-Morgenstern}. $$C^{EFGM}(u, v) = uv(1 + \theta (1 - u)(1 - v)), \quad -1 \le \theta \le 1.$$
    \end{itemize}

    Two dimensional plots of particular copulas from mentioned classes are presented on fig. \ref{fig:copulas}.
    \begin{figure}[!h]
        \centering
        \includegraphics[width=100mm]{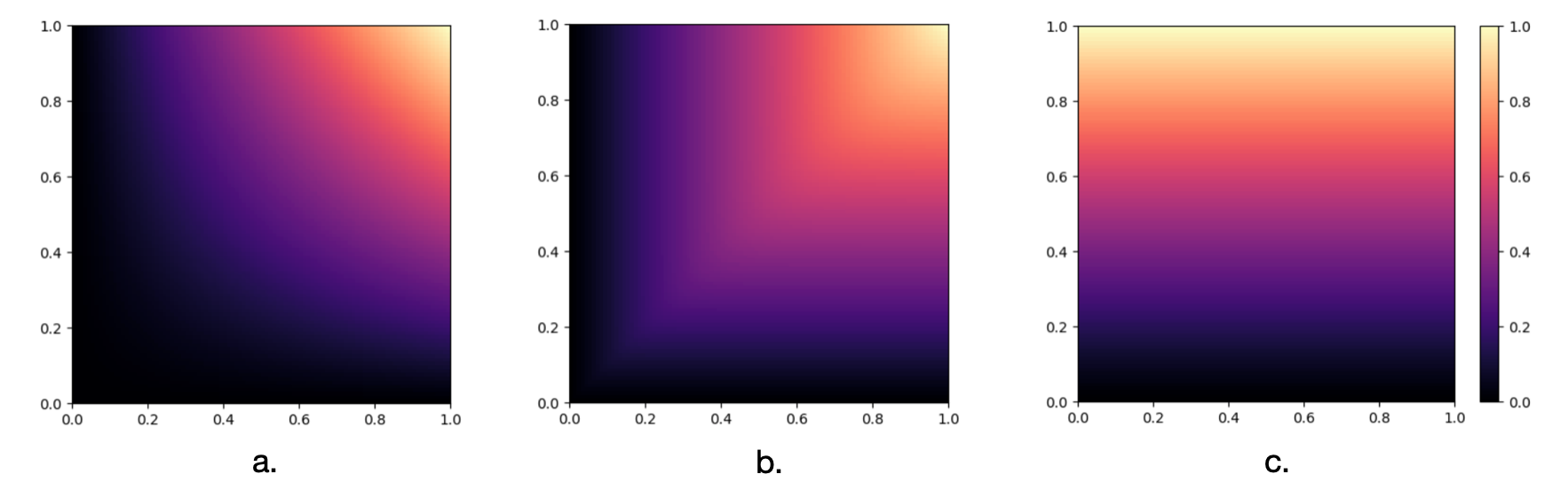}
        \caption{Different copula classes. a. Independent, b. Clayton ($\theta = 10$), c. Gumbel ($\theta=2$)}
        \label{fig:copulas}
    \end{figure}
    
    The classic method for estimating copula parameters is the maximum likelihood method. Assuming that the marginal distribution functions of the random variables $X$ and $Y$ and the joint distribution function have densities, we can calculate the copula density $c(u, v)$ for the random variables $X$ and $Y$:
    $$ c(u, v) = \frac{\partial^2 C(u, v)}{\partial u \partial v} = \frac{f_{X, Y}(x, y)}{f_X(x) f_Y(y)},$$
    where $u = F_X(x)$, $v = F_Y(y)$. 

    The maximum likelihood method can be used to estimate the vector of parameters $\theta = (\beta_1, \beta_2, \alpha)^T$ from samples $X_i = (x_1, \dots, x_n)$ and $Y_i = (y_1, \dots, y_n )$, which are instances of random variables $X$ and $Y$, respectively (in our case, $S^i_t$ and $S^j_t$):

    $$ l(\theta) = \sum^n_{i = 1} \textbf{ln} \: c \left(F^1(x_i; \beta_1), F^2(y_i; \beta_2); \alpha \right) + \textbf{ln} \: f^1(x_i; \beta_1) + \textbf{ln} \: f^2(y_1; \beta_2), $$
    In this model, the maximum likelihood estimate of the parameter vector $\theta$ is:
    $$ \widehat{\theta}_{MLE} = \textbf{argmax}_{\theta} l(\theta). $$

    It is worth noting that this approach to estimating copula parameters is computationally inefficient and the two-step procedure IFM\footnote{Inference for the Margins} is most often used, in which the parameters $\beta_i$ are sequentially estimated, and then the parameter $\alpha$ \cite{estim}.
    
    \item \textbf{Determine underpriced (overpriced) asset}. Knowing the copula of the random vector $(S^i_t, S^j_t)$ of spreads for a pair of selected assets, we can determine the criterion for underpricing (overpricing) of each of the two assets. Following the $4$ property in section \ref{prop}, we calculate the conditional probabilities using the copula:
        $$ h^{1 \mid 2} = \mathbb{P}[U \le u \mid V = v] = \frac{\partial C(u, v)}{\partial v}$$
        $$ h^{2 \mid 1} = \mathbb{P}[V \le v \mid U = u] = \frac{\partial C(u, v)}{\partial u}$$

    Finally, we will say that asset $1$ is underpriced and asset $2$ is overpriced for the level $\varepsilon$ if
    \begin{equation}
        h^{1 \mid 2} < \varepsilon \quad \textbf{and} \quad h^{2 \mid 1} > 1 - \varepsilon,
        \label{eqn:decision1}
    \end{equation}
    and, conversely, that asset $2$ is underpriced and asset $1$ is overpriced if
    \begin{equation}
        h^{1 \mid 2} > 1 - \varepsilon \quad \textbf{and} \quad h^{2 \mid 1} < \varepsilon.
        \label{eqn:decision2}
    \end{equation}
    
\end{enumerate}

\section{Conclusion}
In this work, the methodology for determining whether each of the assets within the trading pair is underpriced(overpriced) is expounded. The shortcomings of the classical method for identifying deviations in the behavior of trading instruments, which utilize the linear correlation coefficient as a measure of divergence, are outlined. An alternative approach for measuring the deviation level in the behavior of financial instruments using copulas is described. The process of constructing a copula for the spread between financial instruments is extensively examined.

Further research anticipates extensive empirical testing of the described method. A pivotal aspect in decision-making involves determining the level $\varepsilon$ (see formulas \ref{eqn:decision1}, \ref{eqn:decision2}). Determining the optimal level of $\varepsilon$ warrants special attention and may be addressed in a separate study, as the relationship between the profitability of the proposed method and the selected level $\varepsilon$ is currently underexplored.

\end{document}